\documentstyle[preprint]{acmconf}
\toappear{ISSAC'98 Poster, 13-15 August 1998, Rostock, Germany}

\newtheorem{th}{Theorem}
\newtheorem{cor}{Corollary}
\newtheorem{prop}{Proposition}
\newtheorem{lemma}{Lemma}
\newtheorem{defi}{Definition}

\newcommand{\qee}{\hfill$\Box$}
\def\r#1{(\ref{#1})}
\def\ci#1{\cite{#1}}
\def\dx{D_x}
\def\dy{D_y}
\def\PX{{\cal P}_x}
\def\PY{{\cal P}_y}
\def\QXY{\Q(x,y)[D_x,\dy]}
\def\QX{\Q(x,y)[D_x]}
\def\QY{\Q(x,y)[D_y]}
\def\QxY{\Q(x,y,\dx)[D_y]}
\def\QyX{\Q(x,y,\dy)[D_x]}
\def\R{|R\}}
\def\L{\{L|}
\def\XM{X_{M_1}}
\def\YM{Y_{M_1}}
\def\lb{\left\{}

\def\1{^{-1}}
\def\proof{{\sl Proof. }}
\def\wt{\widetilde}

\def\ord{{\rm ord}}
\def\lcof{{\rm lcof}}
\def\be{\begin{equation}}
\def\ee#1{\label{#1}\end{equation}}
\def\p{operator}

\def\rg{{\rm rGCD}}
\def\rl{{\rm rLCM}}
\def\lg{{\rm lGCD}}
\def\ll{{\rm lLCM}}
\def\o{\circ}

\def\={\overline }

\def\f{factor}
\def\z{ization}
\def\Q{{\bf Q}}

\begin{document}

\title{Factorization of linear partial differential operators
and Darboux integrability of nonlinear PDEs\thanks{The
research described in this article was partially supported by
INTAS grant 95-IN-RU-412
and Russian Presidential grant 96-15-96834} \\
{\it (ISSAC'98 Poster)}}

\author{S.P.Tsarev \\
    Department of Mathematics\\
Krasnoyarsk State Pedagogical University\\
Lebedevoi, 89 \\
660049, Krasnoyarsk, Russia\\
e-mail: {\tt tsarev@edk.krasnoyarsk.su }\\
}

\maketitle

\abstract{
Using a new definition of generalized divisors we prove that the lattice of
such divisors for a given linear partial differential operator is modular
and obtain analogues of the well-known theorems of the Loewy-Ore theory of
factorization of linear ordinary differential operators. Possible applications
to \f ized Gr\"obner bases computations in the commutative and non-commutative
cases are discussed, an application to finding criterions of Darboux
integrability of nonlinear PDEs is given.
}


\section{ Introduction}
\label{s1}

Factorization is often used for simplification of solution procedures for
polynomials (\f ized Gr\"obner bases computations) and
linear ordinary differential operators (LODO). It is well-known that every
(multivariate)  polynomial factors into product of irreducible polynomials
(in the given
coefficient field) in a unique way; for LODO an analogous result had been
proved by E.Landau \ci{landau} and in a more precise form by A.Loewy
\ci{loewy1,loewy2}: any two different decompositions of a given LODO $L$ into
products of irreducible  LODO $L= P_1 \o\cdots \o P_k =
\=P_1 \o\cdots \o \=P_p$ have the same number of \f s ($k=p$)
and the \f s are pairwise similar  (in some transposed order).
Two (irreducible for simplicity) LODO $L$ and $M$ are called {\em similar}
(or {\em \p s of the same type\/}) if one can find \p s $A$ and $B$ such that
$\ord(A)=\ord(B) < \ord(L)=\ord(M)$ and $A\o L=M\o B$ (see below for more
detail). The theory of \f\z\ of LODO (Loewy-Ore theory) was developed in
\ci{loewy1,loewy2,ore1}. From the algorithmic point of view \f\z\ of LODO was
addressed for the first time in \ci{beke} where an outline of an algorithm for
\f\z\ of LODO with coefficients in the simplest differential field of
rational functions (i.e.\ over $\=\Q(x)$) was given. In the past decade many
improvements of this algorithm and alternative algorithms were proposed (see
\ci{ts96} and references therein), applications to the differential Galois
group computation were given in \ci{singer1}.

Unfortunately very little is known about factorization properties of
linear partial differential \p s (LPDO). The following interesting example was
given by E.Landau (see \ci{blum}): if 
\be
\begin{array}{c}
 P=D_x+xD_y, \quad Q=D_x+1, \\
         R=D_x^2+xD_xD_y+D_x+(2+x)D_y,
\end{array}
\ee{land}
then  $L = Q \circ Q \circ P = R \circ Q$. On the other hand the \p\ $R$ is
absolutely irreducible, i.e.\ one can not factor it into product of first-order
\p s with coefficients in {\em any} extension of $\Q(x,y)$.

This example shows that in order to develop a ``good'' theory of \f\z\ of LPDO 
one shall try to use some  generalization of the notion of a factor (divisor)
for LPDO. Such tricks are very common in the commutative case, as the first
example we may cite Kummer-Dedekind theory of divisors for algebraic number
rings. As proposed by Dedekind we may use ideals of the ring of
algebraic integers of a
given (finite) extension of $\Q$. Since not all ideals in this ring are {\em
principal} ideals (i.e.\ they are not generated as multiples of a {\em single}
 element) we obtain an extension of the notion of a divisor and this suffices
(cf. for example \cite[Ch.12]{ire}) to obtain uniqueness
 of decomposition of any
algebraic integer (i.e.\ of the principal ideal it generates) into product of
prime ``ideal'' divisors.

For the non-commutative ring of LODO (with coefficients in some differential
field, for simplicity 
we will suppose that coefficients belong to $\=\Q(x)$ i.e.
they are rational functions with arbitrary algebraic number coefficients) one
can use the Euclid division algorithm to prove that any left or right
 ideal in this
LODO ring $\Q(x)[\dx]$ is a principal ideal; there are no nontrivial  two-sided
ideals. So there is no possibility (and necessity) of ``ideal'' generalization
of the notion of divisors, the only implication of non-commutativity results in
``similarity'' of factors in different \f\z s of a given LODO.

Another well-known ``unique \f\z'' theorem is the classical Jordan-H\"older
theorem in the theory of finite groups (or finitely generated modules).

In the first half of the XX century a common approach to these (and many more)
cases was proposed. Let us introduce the obvious partial order in the set of
(left) ideals: $I_1 \leq I_2$ if $I_1 \supset I_2$; we  call $I_1$ a divisor of
$I_2$ in such a case. Then instead of \f\z s 
\be
L=L_1\o \ldots \o L_k
\ee{fac}
of an element $L$ of the ring we will consider chains
$  |L\rangle >|L_2\o \ldots \o L_k\rangle >
|L_3\o \ldots \o L_k\rangle > \ldots > |L_k\rangle
 > 0 = |1\rangle
$ of corresponding (left) principal ideals. Irreducibility of factors
corresponds to {\em maximality} of this chain, i.e.\ impossibility to insert
intermediate ideals between any two elements of the chain.
 The partially ordered
set $\cal M$ (also called poset) of ideals in the cited above ``good'' cases
has the following two fundamental properties:

a) for any two elements $A,B \in \cal M$ one can find a unique $C=\sup(A,B)$,
i.e.\ such $C$ that $C \geq A$, $C \geq B$, and $\forall X \in \cal M$,
$(X \geq A , X \geq B) \Rightarrow X \geq C$. Analogously there exist
a unique $D=\inf(A,B)$,
 $D \leq A$, $D \leq B$, $\forall X \in \cal M$,
$(X \leq A , X \leq B) \Rightarrow X \leq D$.
Such posets are called {\em lattices}. $\sup(A,B)$ and $\inf(A,B)$
 correspond to the least common multiple and the greatest common divisor
for the cases of number rings and LODO. In our cases a lattice will always have
{\em zero} i.e.\ an element $0 \in \cal M$ such that  $\forall X \in \cal M$,
$ X \geq 0$. For simplicity (and following the established tradition)
$\sup(A,B)$ will be hereafter denoted as $A+B$ and $\inf(A,B)$ --- as
$A\cdot B$;

b) For any three $A,B,C \in \cal M$ the following {\em modular identity} holds:
\be
(A\cdot C + B)\cdot C = A\cdot C + B\cdot C
\ee{mo}

This weaker form of distributivity was discovered by Dedekind. The theory of
modular lattices (i.e.\ posets with the above two properties, such posets are
also called ``Dedekind structures'') has beautiful (for our purpose :--)
results. Namely these two simple properties are sufficient to  prove
the following four elegant theorems (cf. \ci{bir,grat,jac1}):
\begin{th}
\label{T1}
 { \rm (Jordan-H\"older-Dedekind chain condition)} Any two finite maximal
chains 
\be
L  > L_1 > \cdots > L_k > 0
\ee{chain}
$$
L  > M_1 > \cdots > M_r > 0
$$
for a given $L \in \cal M$ have equal length: $k=r$.
\end{th}
\begin{th}
\label{T2}
{(\rm Kurosh \& Ore)} If $L= L_1 +L_2+\ldots +L_p = 
M_1 +M_2+\ldots +M_r$ are two noncancellable $\sup$-representations of $L \in
\cal M$ then $p=r$ and for every $L_i$ one can find $M_j$ such that
$L= L_1 +\ldots +L_{i-1}+M_j+L_{i+1}+\ldots +L_p$.
\end{th}
We recall that a $\sup$-representation
\be
 L= L_1 +L_2+\ldots +L_k
\ee{sup}
is called non-cancellable if $\forall i$,
$(L_1 +\ldots +L_{i-1}+L_{i+1}+\ldots +L_k) \neq L$
and each $L_i$ can not be $\sup$-represented: $L_i \neq A_i + B_i$
for $A_i \neq L_i$, $B_i \neq L_i$.

We call a $\sup$-representation \r{sup} a {\em direct sum} if
the set $\{L_i\}$ is {\em independent} that is 
$\forall i$,
$(L_1 +\ldots +L_{i-1}+L_{i+1}+\ldots +L_k) \cdot L_i =0$ and
each $L_i$ can not be represented as a  sum of two independent elements.
Direct sums are denoted $L= L_1 \oplus L_2 \oplus \ldots \oplus
L_k$. An element $A \in \cal M$ is called indecomposable if
$A \neq B \oplus C$,  $B \neq 0$, $C \neq 0$.
\begin{th}
\label{T3}
{\rm (O.Ore)} Let an element $L$ of a modular lattice have finite maximal
chains \r{chain} and $L= L_1 \oplus  \ldots \oplus
L_k =  M_1 \oplus  \ldots \oplus M_r$ with indecomposable $L_i$, $M_j$.
Then $k=r$ and  $\forall L_i$ one can find $M_j$ such that
$L= L_1 \oplus\ldots 
\oplus L_{i-1}\oplus M_j\oplus L_{i+1}\oplus\ldots \oplus L_p$.
\end{th}
Let us call $l(L):=k+1$ the {\em length} of $L \in \cal M$
if $L$ has a finite maximal chain \r{chain} (of length $k+1$). We
set $l(0)=0$. The length of a LODO is equal to the number of irreducible
factors in decomposition \r{fac}.
\begin{th}
\label{T4}
If all elements of a modular $\cal M$ have finite length then
$ l(A+B)+l(A\cdot B)= l(A)+l(B)$.
\end{th}
These theorems give a unified approach to many well known facts in the theory
of groups (and group representations), commutative and non-commutative rings;
in particular they encompass many results of the Loewy-Ore theory of \f\z\ of
LODO.

Let us prove here for completeness  that the poset of (left) ideals of a
(non-commutative) ring is a modular lattice. Firstly we notice that
$\sup(A,B)=A+B$ corresponds to the intersection of ideals $A$,$B$;
$\inf(A,B)=A\cdot B$ corresponds to the ideal composed of the sums
$a+b$, $a \in A$, $b \in B$. Then if $x \in A\cdot C +B\cdot C$ (the r.h.s of
\r{mo}) then $x \in C$, $x= a+b$, $a \in A$, $a \in C$, $b \in B$, $b \in C$.
Obviously $a+b \in C$, $a+b \in A\cdot C +B$, so $A\cdot C +B\cdot C
\subset (A\cdot C +B)\cdot C$. Vice versa if $x \in (A\cdot C +B)\cdot C$
then $x \in C$, $x= a+b$, $a \in A$, $a \in C$, $b \in B$ $\Rightarrow$
$b=x-a \in C$ so $b \in B \cdot C$ and $x \in A\cdot C +B\cdot C$
which proves  $(A\cdot C +B)\cdot C
\subset A\cdot C +B\cdot C$.

The basic notion of similarity also exists for modular lattices:
\begin{defi}
\label{def1}
Two elements $A$, $B$ 
of a modular lattice $\cal M$ are called similar if one can find
$C \in \cal M$ such that $A\cdot C=B\cdot C=0$ and $A+C=B+C$
(i.e.\ $A\oplus C= B\oplus C$).
\end{defi}
We will need also the notion of similarity of {\em intervals} or {\em
quotients} $[B/A] := \{ X \in {\cal M} | A \leq X \leq B\}$ for pairs $A \leq B$.
\begin{prop}
\label{prop1}
If $A$, $B$ are elements of a modular lattice $\cal M$ then the intervals
$I_1= [A/(A\cdot B)]$ and $I_2=[(A+B)/B]$ are {\em projective}, $I_1 \sim I_2$,
i.e.\ isomorphic with specific poset 
isomorphisms $\phi : I_1 \rightarrow I_2$,
$\phi(X)=X+B$, $\psi(X)=A\cdot X$, $\psi = \phi^{-1}$.
\end{prop}
\begin{defi}
\label{def2}
Two intervals $[B_1/A_1]$, $[B_2/A_2]$ are called similar if there exists a
finite sequence of projective intervals $[B_1/A_1] \sim I_1 \sim I_2 \sim
\ldots \sim I_k \sim [B_2/A_2]$.
\end{defi}
One can prove that in Theorems~\ref{T1}--\ref{T3} the corresponding factors
(intervals) are similar (in some transposed order). Similarity of intervals
in Theorem~\ref{T1} gives similarity of the respective irreducible factors in
\r{fac} or isomorphism of the  factorgroups (factormodules) for the modular
lattice of normal subgroups of a given (finite)  group (resp. submodules).

The case of the ring of LPDO is more complicated. It has no two-sided ideals
and left (right) ideals are no longer principal ideals in the general case.
Certainly the poset of {\em all} left (right) ideals is a modular lattice.
But unfortunately we can not use the above results: for a LPDO $L$ we get
finite chains \r{chain} of left ideals (the ring of LPDO is Noetherian,
see \ci{bjork}) but the intervals in any chain are not (as a rule)
``irreducible'' i.e.\ one can always insert intermediate ideals between some
of them so the length of chains \r{chain} for a given $L$ is not bounded. For
example for arbitrary LODO $L \in \Q(x)[\dx] \subset \QXY$ we can take
$|L\rangle > |L\rangle+|\dy^m\rangle >|L\rangle+|\dy^{m-1}\rangle >
\ldots |L\rangle+|\dy\rangle >0=\QXY$. Even the simplest $\dx$ becomes
``reducible''! Similar  infinite examples exist for decompositions into
(direct) $\sup$-sums. So Theorems~\ref{T1}--\ref{T4} are useless.

We conclude that the poset of {\em all} (left) ideals of LPDO is too
``large''. For the commutative case of multivariate polynomials one can limit
oneself to principal ideals and get the desired modular lattice with finite
chains. Again for LPDO the poset of (left) principal ideals is too
``small'': it does not form even a lattice. For example for the two operators
$P$, $Q$ in \r{land} the intersection of the left principal ideals
$|P\rangle \bigcap |Q\rangle$ (their ``LCM'') is no longer principal: 
one can easily check directly that there are no second-order common left
multiples of both $P$, $Q$ but we have {\em two} linearly
 independent third-order operators divisible by $P$, $Q$:
$$
L_{31}= \big(x\dx\dy + (x-1)\dy -\dx -1\big)\o P=
$$
$$
\quad  \big(x^2\dy^2 + x\dx\dy -(x+1)\dy - \dx\big) \o Q,
$$
$$
L_{32}= \big(\dx^2  + 2\dx +1\big)\o P= Q\o Q \o P =
$$
$$
 \quad \big(\dx^2 + x\dx\dy +(x+2)\dy + \dx\big) \o Q = R \o Q,
$$
so there is no ``least'' common left multiple. Analogously we can directly
check that these $L_{31}$, $L_{32}$ have only $Q$, $P$ as their common
right divisors so $L_{31}$, $L_{32}$ have no ``greatest'' right common divisor.
Also as the E.Landau's example shows the Jordan-H\"older-Dedekind chain
condition fails for principal ideals.

Below (section~\ref{s2}) we define an ``intermediate'' poset of
 ``codimension~1'' left ideals which is larger than the poset of principal left
ideals but smaller than the lattice of all left ideals. This new poset 
of ``generalized divisors'' provides
all the necessary properties: it is a modular lattice with finite maximal
chains \r{chain} for every element and finite decompositions into (direct)
$\sup$-sums so the basic Theorems~\ref{T1}--\ref{T4} are applicable;
any first-order LPDO is irreducible and any LODO $L$ irreducible in
$\Q(x)[\dx]$ remains irreducible (as LPDO) in our poset.
This is our main result.

For applications the most important property of our modular lattice
of generalized divisors would be certainly the possibility to decompose
operators into $\sup$-sums in an overdetermined system of LPDO
\be
\lb
\begin{array}{c}
L_1f=0,\\
L_2f=0,\\
\cdots \\
L_kf=0.\\
\end{array}
\right.
\ee{sy1}
Suppose that $L_1=A_1 + \ldots A_p$ for some left ideal divisors then since
each $A_i$ is finitely generated (see \ci{bjork}): $A_i=|L_{i1},\ldots,
L_{is_i}\rangle$,  we can decompose \r{sy1} into union of systems
\be
\lb
\begin{array}{c}
L_{11}f=0,\\
\cdots \\
L_{1s_1}f=0,\\
L_2f=0,\\
\cdots \\
L_kf=0.\\
\end{array}
\right.
\quad \cdots \quad
\lb   \begin{array}{c}
L_{p1}f=0,\\
\cdots \\
L_{ps_p}f=0,\\
L_2f=0,\\
\cdots \\
L_kf=0.\\
\end{array}
\right.
\ee{sy2}
Sums of solutions of \r{sy2} are obviously solutions of \r{sy1} and we
conjecture that they span the whole space of solutions of \r{sy1}.
Also we need an algorithm for such $\sup$-decompositions of LPDO
(see section~\ref{conc} for the discussion).

Substitution of \r{sy1} with \r{sy2} is an analogue of the well-known
factorization technique for commutative Gr\"obner bases computations.
This technique considerably reduces the complexity of computations in many
practical cases. The overdetermined systems of type \r{sy1} with one or many
unknown functions are typical in many
applications (\ci{olver,schwarz}): computation of conservation laws,
symmetries and invariant solutions of systems and single nonlinear
ODEs and PDEs. For any system \r{sy1} one may use the standard Janet-Riquier
technique (\ci{jan,reid,riq,schwarz}) of reduction of \r{sy1} to the so called
passive (standard, normal) form. In the case of constant coefficient systems
\r{sy1} (in the commutative case) this algorithm practically coincides with the
Gr\"obner algorithm (for total degree+weight ordering). Unfortunately the
complexity of Janet-Riquier algorithm is very high even for modest LPDO
systems. Recently  one interesting contribution to reduction of the
complexity for computation of the genus (roughly speaking this is the
``dimension'' of the solution space) of \r{sy1} was given in \ci{grig2}.
Our approach may help in decomposition of the solution space of \r{sy1}
into ``irreducible submanifolds''. Further possible generalizations and
applications to the commutative case are discussed in section~\ref{conc}.

Another connection of our definition of factorization of LPDO
and integrability properties of nonlinear PDEs is discussed in
 section~\ref{di}:
the established in \ci{anderson,juras,sokolov}
 criterion of Darboux integrability \ci{darboux}
("explicit" integrability) of such nonlinear PDEs is
equivalent to generalized factorization of the corresponding
linearized equation. This gives a new insight into possible generalizations
of the notion of Darboux integrability of higher-order
nonlinear PDEs which is now under investigation.

\section{ Divisor ideals of LPDO}
\label{s2}

We study general LPDO
\be
L=\sum_{|\vec i|\leq m} a_{i_1\cdots i_n}(\vec x)
D_{x_1}^{i_1}D_{x_2}^{i_2}\cdots D_{x_n}^{i_n},
\ee{lpdo}
$|\vec i| = i_1 + \ldots +i_n$, $\vec x = (x_1, \ldots, x_n)$,
$D_{x_i}=\partial/\partial x_i$, $\ord(L):=m$.
For simplicity and without loss of generality we will suppose that the number
of the independent variables is $n=2$, $x:=x_1$, $y:=x_2$, and the coefficients
$a_{ij}(x,y)$ in \r{lpdo} are rational functions with rational coefficients,
$a_{ij}(x,y) \in \Q(x,y)$, so $L \in \QXY$.

It is straightforward to check that for every finite set of LPDO
$L_1$, \ldots, $L_k$ one may algorithmically find all their common left 
multiples (left c.m.) up to fixed order $N$:
take $M_1\o L_1 =\ldots = M_k\o L_k$ with $\ord(M_i)=N-\ord(L_i)$ and
indefinite coefficients, then we get a linear algebraic (not differential!)
system for the coefficients of $M_i$; the number of equations in this linear
system will be less than the number of the unknown coefficients for
sufficiently large $N$, so the set of left (right) c.m.\ is always
nonempty.

All these and subsequent results are 
certainly invariant w.r.t.\ substitution of
left ideal with right ideals; application of the usual  adjoint operation will
suffice for this purpose. We will denote the left (right) principal ideal
generated by LPDO $L$ with $|L\rangle$ (resp. $\langle L|$).
\begin{defi}
\label{def11}
The left LPDO ideal $\ll(|L_1\rangle, \ldots , |L_k\rangle):=
|L_1\rangle \bigcap \ldots  \bigcap  |L_k\rangle$ is called the left least
common multiple of LPDO $L_i$.
\end{defi}
This $\ll$ is always non-empty and (see Introduction) not principal in the
general case.
\begin{defi}
\label{def12}
The same ideal
$|L_1\rangle \bigcap \ldots  \bigcap  |L_k\rangle$ will be also called
 the left greatest common divisor of  $L_i$.
\end{defi}
{\bf Remark}. This is serious :--) Below the reader will see that this is the
key to the whole trick.
\begin{defi}
\label{def13}
We call two LPDO $L$, $R$ a (generalized) divisor \p\ couple for LPDO $M$
if there exist LPDO $X$,$Y$, $Q$ such that
\be
\begin{array}{c}
X\o M= Y\o R, \\
X\o L= Y\o Q.
\end{array}
\ee{ddd}
\end{defi}
One may informally think that $M \cong L\o R$ and if in fact $M$ factors into
the product of $L$, $R$ then we may choose $X=Q=1$, $Y=L$ in \r{ddd}.

A divisor \p\ couple is called {\em nontrivial} if $\ord(L)>0$, $\ord(R)>0$ and
$L$, $R $ are {\em not} divisible by $M$, i.e.
$L \neq M\o P$, $R \neq K\o M$. In this case we will say that
the \p s $M$ and $R$
($M$ and $L$) have nontrivial (generalized) right (resp. left) common divisor.

{\bf Remark}. These definitions actually say that we can restore the (greatest)
common divisor if we can find (least) common multiples; for the case of LODO if
$Z=\ll(M,R)= X\o M=Y\o R$ and $V=\rl(X,Y)=X\o L=Y\o Q$ then $M=L\o G$, $R=Q\o
G$, $G=\rg(M,R)$. This explains Definition~\ref{def12}. For integral domains
(commutative rings with $1$ and no zero divisors) if two elements have LCM
(i.e.\ some common multiple such that this c.m.\ divides any other c.m.)  then
they automatically have GCD
(i.e.\ some common divisor such that this divisor is divided by any other common
divisor); the converse is not true in general; also $LCM(ac,bc)=c\cdot
LCM(a,b)$ when one of them exist, this is again not true for GCD (the author
thanks Dr.~N.N.Osipov who communicated to him these facts for integral
domains).
\begin{lemma}
\label{le1}
If \r{ddd} holds and for some $X_1$, $Y_1$ we have $X_1\o M=Y_1\o R$, then
$X_1\o L=Y_1\o Q$.
\end{lemma}
\proof Let us find some left c.m.\ of $X_1$, $X$:
$\wt X=\=X\o X_1 =\=X_1 \o X$. Then $\=X\o X_1 \o M = \=X\o Y_1\o R
= \=X_1\o X\o M = \=X_1 \o Y \o R$ so (since the ring of LPDO has no zero
divisors) $\=X\o Y_1= \=X_1\o Y$ and $\=X_1\o  Y \o Q = \=X_1 \o X \o L
=\=X \o X_1 \o L = \=X\o Y_1 \o Q$ hence $X_1 \o L = Y_1 \o Q$. \qee

So \r{ddd} does not depend on the choice of the $\ll(M,R)=X\o M= Y\o R$.
\begin{lemma}
\label{le2}
If \r{ddd} holds and some right c.m.\ of $M$, $L$ is chosen $M\o \=X=L\o \=Y$
then
\be
\begin{array}{c}
M\o \=X=  L \o \=Y,\\
R\o \=X = Q \o \=Y.
\end{array}
\ee{dd2}
\end{lemma}
\proof $Y\o Q\o \=Y = X \o L \o \=Y = X\o M \o \=X = Y \o R \o \=X$
$\Rightarrow$ $Q\o \=Y = R \o \=X$. \qee

From \r{ddd} and Lemma~\ref{le1} we conclude that the set of \p s $L$ forming a
generalized divisor couples with fixed $R$, $M$ is a right ideal $\L$;
from \r{dd2} we see that for fixed $M$, $L$ \p s $R$ form a left ideal
$\R$.
\begin{defi}
\label{dic}
Left ideal $\R$ and right ideal $\L$ form a (generalized) divisor ideal couple
for an \p\ $M$ (we denote this fact as $\L M \R$) if:\\
a) any $R \in \R$, $L \in \L$ form a divisor \p\ couple for $M$ i.e.\ 
   \r{ddd} holds;\\
b) if some LPDO $L$ forms divisor \p\ couples for $M$ with {\em every}
$R \in \R$ then $L \in \L$;\\
c) if some LPDO $R$ forms divisor \p\ couples for $M$ with {\em every}
$L \in \L$ then $R \in \R$.
\end{defi}
\begin{lemma}
\label{le3}
Let $\L M\R$ be a divisor ideal couple for $M$. Then for every $M_1 \in \R$
we can find a unique right ideal $\{Q_1|$ such that $\{Q_1|M_1\R$.
\end{lemma}
\proof Since $M_1 \in \R$ then for every $L \in \L$ we have the unique
(Lemma~\ref{le1}) $Q_1$ such that $\exists X_1,Y_1$,
\be
\begin{array}{c}
X_1\o M= Y_1\o M_1, \\
X_1\o L= Y_1\o Q_1.
\end{array}
\ee{d3}
Take some left c.m.\ of $M_1$, $R$ for $R \in \R$:
$Z_1 = \XM \o M_1 = \YM \o R$, and some left c.m.\ of $Y_1$, $\XM$:
$Z_2 = \=Y_1\o \XM = \=\XM\o Y_1$. Then
$\=\XM\o X_1 \o M = \=\XM \o Y_1 \o M_1 = \=Y_1 \o \XM \o M_1 = 
\=Y_1 \o \YM \o R$, so we get $\=\XM\o X_1 \o M = \=Y_1 \o \YM \o R$.
Using Lemma~\ref{le1} we conclude
$\=\XM\o X_1 \o L = \=Y_1 \o \YM \o Q$
for any $L \in \L$ and the corresponding $Q$.
Consequently $\=Y_1\o \YM \o Q = \=\XM \o X_1 \o L = \=\XM \o Y_1 \o Q_1 =
\=Y_1 \o \XM \o Q_1$. Cancelling $\=Y_1$ we get $\YM \o Q = \XM \o Q_1$ and
finally
\be
\begin{array}{c}
\XM\o M_1= \YM\o R, \\
\XM\o Q_1= \YM\o Q,
\end{array}
\ee{d4}
which shows that any $Q_1$ in \r{d3} (it depends on $L \in \L$) form
a divisor \p\ couple for $M_1$ with any $R \in \R$ so the condition a)
of Definition~\ref{dic} holds.
In order to prove the condition c) of it for $\{Q_1|M_1\R$ we suppose \r{d4} to
be true for some LPDO $R$ and all $Q_1$ obtained from \r{d3} with
$L \in \L$. Fixing $\XM$, $\YM$, $X_1$, $Y_1$ and using the same definition of
$\=\XM$, $\=\YM$: $Z_2 = \=Y_1\o \XM = \=\XM\o Y_1$ we get
$\=Y_1 \o \XM \o M_1 = \=\XM \o Y_1 \o M_1 = \=Y_1 \o \YM \o R =
\=\XM \o X_1 \o M$,
$\=Y_1\o \XM \o Q_1 = \=Y_1 \o \YM \o Q = \=\XM \o Y_1 \o Q_1  = \=\XM
\o X_1 \o L$ so
\be
\begin{array}{c}
\=\XM\o X_1\o M= \=Y_1 \o\YM\o R, \\
\=\XM\o X_1 \o L= \=Y_1 \o\YM\o Q,
\end{array}
\ee{xym}
that is \r{ddd} holds (Lemma~\ref{le1})
for this $R$ and all $L \in \L$ so $R \in \R$. We prove the condition b) of
Definition~\ref{dic} in 
the same way: if \r{d4} holds for some LPDO $Q_1$ and all
$R \in \R$ ($Q$, $\XM$, $\YM$ depend
 on $R$) we take $R=M \in \R$ so \r{d4} becomes
$$
\begin{array}{c}
\wt \XM \o M_1= \wt \YM\o M, \\
\wt \XM \o Q_1= \wt \YM\o \wt Q.
\end{array}
$$
In this case we see from \r{d3}
that actually we may set (Lemma~\ref{le1}) $\wt \XM=Y_1$, $\wt \YM =X_1$,
$L:=\wt Q$. So we have \r{d3}, \r{d4} with $R \in \R$ and we shall
 prove \r{ddd} for the constructed $L$, $Q$, $M$, $R$.
Again $Z_2 = \=Y_1\o \XM = \=\XM\o Y_1$, 
$\=\XM \o Y_1 \o M_1 = \=\XM \o X_1 \o M = 
\=Y_1 \o \XM \o M_1 = \=Y_1 \o \YM \o R$,
$ \=\XM\o X_1 \o L =  \=\XM \o Y_1 \o Q_1 =
\=Y_1\o \XM \o Q_1 = \=Y_1 \o \YM \o Q $,
i.e.\ we have \r{xym} again or (Lemma~\ref{le1}) we have \r{ddd}.
 \qee

Lemma~\ref{le3} essentially says that 
the right parts of $\L M\R$ are internally
characterizable as some special left ideals. The same is true for $\L$. We will
call such $\R$ right divisor ideals 
or r.d.i.\ (they are left ideals of the ring
of LPDO :--) and $\L$ --- left divisor ideals or l.d.i.
(they are right ideals). Any principal left ideal is a r.d.i.:
$|R_0\rangle = |R_0\}$ since we obviously have $\{1|R_0|R_0\}$, that is
\r{ddd} with $R=P\o R_0$, $M= R_0$, $Y=1$, $X=P$, $Q=P \o L$,
($P$, $L$ being arbitrary LPDO).

A divisor ideal couple $\L M\R$ is called {\em trivial} if either
$\L =\langle M|$ (then $\R =0 = |1\rangle$) or
$\R = |M\rangle$, $\L =0$.

On the other hand as we will see in the next section there are divisor ideals
which are not principal and not every (left) ideal is r.d.i.

Let us now prove that the set of r.d.i.\ with the natural ordering
($|R_1\} \geq |R_2\}$ iff $|R_1\} \subset |R_2\}$)
forms a lattice. Namely for two r.d.i.\ $|R_1\}$, $|R_2\}$
we take their intersection as their lLCM:
$\ll(|R_1\}, |R_2\}) := \sup(|R_1\}, |R_2\}) := |R_1\} + |R_2\}:=
|R_1\}\bigcap |R_2\}$. Then for $M \in |R_1\}\bigcap |R_2\}$ we find two
corresponding l.d.i.\ (Lemma~\ref{le3}): $\{L_1|M|R_1\}$,
$\{L_2|M|R_2\}$. Now let us take {\em all} LPDO $L$ such that \r{ddd} holds for
every $R \in |R_1\}\bigcap |R_2\}$. The right ideal $\L$ of such \p s forms the
divisor ideal couple with $\R = |R_1\}\bigcap |R_2\}$ for $M$ since a) and b)
in Definition~\ref{dic} hold automatically  and $\L \supset 
\{L_1|\bigcup \{L_2|$ so every \p\ $R$ such that \r{ddd} holds for all $L \in
\L$ forms a divisor \p\ couple with all $L\in \{L_1|$ and
$L\in \{L_2|$ so $R \in |R_1\}\bigcap |R_2\}$ and the condition c) holds.
As we will see in the next section the constructed $\L$ is in general greater
than the set of all sums of elements of $\{L_1|$ and $\{L_2|$.
Obviously this $\L$ plays the role of $\lg(\{L_1|,\{L_2|)\equiv 
\{L_1|\cdot\{L_2|$ in the poset of all l.d.i.\ The determination of 
$\R =\rg(|R_1\}, |R_2\}) := \inf(|R_1\}, |R_2\}) := |R_1\} \cdot |R_2\}$ is
obtained in the same way: we now take $\L = \rl(\{L_1|,\{L_2|) :=
\{L_1| \bigcap \{L_2|$ and the corresponding 
$\R \supset |R_1\}\bigcup |R_2\}$ is defined using \r{ddd}.
So our lattice of r.d.i.\ does not form a sublattice of the lattice of all right
ideals of LPDO, it changes $\inf$; such subsets are called
``meet-sublattices''.

\section{Coordinatization of divisor ideals}
\label{s3}

Any (non-commutative) ring $R$ satisfying the so called Ore condition
(absence of zero divisors and
existence of at least one common multiple for every two non-zero elements)
may be imbedded into a skew field (non-commutative ring with division) built
with formal quotients $L\1 \o M = \=M\o \=L\1$, $L,M,\=L,\=M
 \in R$ (\ci{ore2}).
Let us take $R=\QX$ and the corresponding skew field $\Q(x,y,\dx)$.
We can form a ring $\QxY$ of \p s of the type
$L= a_0\dy^n + a_1 \dy^{n-1} + \ldots + a_n$, $a_i = (L_i)\1\o M_i \in
\Q(x,y,\dx)$ if we define the corresponding $\dy$-differentiation for the
coefficients: \\
$\partial\big((L_i)\1\o M_i\big)/ \partial y =
(L_i)\1\o \partial (M_i)/\partial y-(L_i)\1\o L_i' \o (L_i)\1 $
with $\partial (M_i)/\partial y = \partial \big(
m_0(x,y)\dx^k + \ldots + m_k(x,y)\big)/\partial y =$ \\
$(\partial m_0(x,y)/ \partial y) \dx^k + \ldots + \partial m_k(x,y)/\partial y$;
$L_i' \in \QX$ is defined via $\dy \o L_i = L_i \o \dy + L_i'$ $\Rightarrow$
$L_i\1 \o \dy = \dy \o L_i\1 + L_i \1 \o L_i' \o L_i\1$.
As explained in \ci{ore1} the basic facts of the Loewy-Ore theory (namely the
existence of the Euclid division algorithm, GCDs, LCMs) hold also for
\p s with coefficients in differential skew fields. Any (left) ideal is again
a principal ideal, they form a modular lattice.

We have the natural projections $\PX: \QXY \rightarrow \QyX$
and $\PY: \QXY \rightarrow \QxY$. For {\em any} left 
ideal $I \subset \QXY$ we obtain two principal ideals $\PX(I)= |I_x\rangle$,
 $\PY(I)= |I_y\rangle$, their generators $I_x$, $I_y$ are called
{\em coordinates} of $I$, $I_x \in \QyX$, $I_y \in \QxY$, we also normalize
them $\lcof_{\dx}(I_x)=1$, $\lcof_{\dy}(I_y)=1$.
Strictly speaking $\PX(I)$ and $\PY(I)$ are not ideals in the respective rings;
in order to make them ideals we need to multiply all their element by
various $L\1$, $L \in \QX$ (resp.\ $L \in \QY$).
 This will be implicitly done hereafter.

First we remark that if we have $\L M \R$ then for the coordinates
\be
\PX(M)= L_x\o R_x
\ee{px}
which gives a heuristic foundation for the definitions of the previous section.
Also obviously $\PX(\ll(|R_1\},|R_2\})) =
\ll(\PX(|R_1\}),\PX(|R_2\}))$ which in turn gives \\
 $\PX(\lg(\{L_1|,\{L_2|)) =
\lg(\PX(\{L_1|),\PX(\{L_2|))$ due to \r{px}. Symmetrically \\
 $\PX(\rg(|R_1\},|R_2\})) =
\rg(\PX(|R_1\}),\PX(|R_2\}))$, \\
 $\PX(\rl(\{L_1|,\{L_2|)) =
\rl(\PX(\{L_1|),\PX(\{L_2|))$.

The following lemma plays the key role in the subsequent proofs.
\begin{lemma}
\label{le5}
If a r.d.i.\ $\R \subset \QXY$ contains two elements $A\o P$, $B \o P$
such that $A \in \QX$, $B \in \QY$ then $P \in \R$.
\end{lemma}
\proof Using  \r{dd2} we obtain $A \o P \o \=X = Q \o \=Y$,
$B \o P \o \=X = S \o \=Y$ for some $S$, $Q$. Then $P \o \=X =
A\1 \o Q \o \=Y \in \QxY$, $B \o P \o \=X = B \o A\1 \o Q \o \=Y
= S \o \=Y$ $\Rightarrow$ 
\be
 B \o A\1 \o Q = S \in \QXY,
\ee{ul5}
$B\in \QX$. We will prove that in such circumstances $Q$ is divisible by
$A$: $Q=A\o \=Q$, $\=Q \in \QXY$. Without loss of generality we may suppose
$\lcof(B)=1$ so \r{ul5} reads
\be
\begin{array}{l}
\underbrace{\left(
\dy^m + b_1(x,y)\dy^{m-1} + \ldots + b_m(x,y) \right)}_B \o {} \\[5em]
\quad \underbrace{\left( C_0 \dy^n + C_1 \dy ^{n-1} + \ldots 
+ C_n\right)}_{A\1\o Q} =S,
\end{array}
\ee{ul51}
with $C_i \in \Q(x,y,\dy)$. The leading coefficient of the l.h.s.
(in $\QxY$)
of \r{ul51} is $C_0$ so since $S \in \QXY$, $C_0 \in \QX$.
Then the coefficient of $\dy^{m+n-1}$ in \r{ul51} will be
$C_1+\partial C_0 /\partial y + b_1 C_0 \in \QX$ so also $C_1 \in \QX$.
Using induction we get $C_i \in \QX$ so $A\1\o Q = \=Q \in \QXY$.
This gives us the possibility to cancel $A$ in $A \o \=X = Q \o \=Y$
obtaining $\=X= \=Q \o \=Y$. Then the \p\ $P \in \R$ since $P\o \=X =
(P\o \=Q) \o \=Y$ in \r{dd2}. \qee

{\bf Remark}. Actually we used only the fact $\lcof_{\dy}(B) =1$.
\begin{cor}
\label{cor1}
If a r.d.i.\ $\R$ contains two elements $A$, $B$, $A \in \QX$, $B \in \QY$
then $\R$ is trivial, $\R =0 = |1\rangle = \QXY$.
\end{cor}
This Corollary explains why r.d.i.\ are ``codimension 1'' ideals:
if we will take a ``codimension 2'' left ideal generated by $\dx^n$, $\dy^m$,
(solutions of the corresponding system $\dx^nf=0$, $\dy^mf =0$ are
functions of 2 variables parameterized by several constants and not by
functions of 1 variable), it is contained only in the trivial r.d.i.\ $|1\}$.
The set of divisor ideals is a bit larger than the set of principal ideals
(which are obviously 
``codimension 1'' ideals --- solutions of the 
 corresponding system of 1 equation are parameterized by 
functions of 1 variable).
\begin{th}
\label{T5}
Two r.d.i.\ $|R_1\}$, $|R_2\}$, coincide iff their coordinates coincide i.e.
iff $\PX(|R_1\})=\PX(|R_2\})$, $\PY(|R_1\})=\PY(|R_2\})$.
\end{th}
\proof Let some $R \in |R_1\}$. 
Since  $\PX(|R_1\})=\PX(|R_2\})$ we may find $\=R \in |R_2\})$, such that $C\o
R = \=R$, $C = (C_1)\1 C_2 \in \Q(x,y,\dy)$. Multiplying with $C_1$ we get
$C_2 \o R = C_1 \o \=R = \wt R \in |R_2\}$. Hence for some $C_2 \in
\QY$ we have $C_2 \o R \in |R_2\}$. Analogous consideration w.r.t.
$\PY$ give  $K_2 \o R \in |R_2\}$ for $K_2 \in \QX$, thus
$R \in |R_2\}$ (Lemma~\ref{le5}) hence $|R_1\} \subset |R_2\}$.
Symmetrically $|R_2\} \subset |R_1\}$. \qee

{\bf Remark}. This is obviously not true for arbitrary ideals:
the ideal generated by $\dx$, $\dy$ have the same projections
as the trivial $|1\rangle$.
\begin{cor}
\label{cor2}
The lattice of r.d.i.\ (l.d.i.) is modular.
\end{cor}
\proof Since the modular identity \r{mo} holds for projections
due to modularity of the lattice of the (principal) left ideals of
$\QxY$, $\QyX$, and GCD, LCM are preserved after projections,
we conclude $\PX((A\cdot C + B)\cdot C) = \PX(A\cdot C + B\cdot C)$,
 $\PY((A\cdot C + B)\cdot C) = \PY(A\cdot C + B\cdot C)$,
so $(A\cdot C + B)\cdot C = A\cdot C + B\cdot C$. \qee
\begin{cor}
\label{cor3}
For any r.d.i.\ $\R$ the length $k+1$ of a chain of r.d.i.
$\R > |R_1\} >  \ldots > |R_k\} > 0$ is limited: $k+1 \leq \ord_{\dx}(\R_x)+
\ord_{\dy}(\R_y)$.
\end{cor}
\proof Since $\PX(|R_i\})$, $\PY(|R_i\})$ give chains of divisors
of $\PX(|R\})$, $\PY(|R\})$ in $\QyX$ (resp. $\QxY$), the adjacent elements of
 $\PX$-projected chain may differ only in $\ord_{\dx}(\R_x)$ places
(resp. in $\ord_{\dy}(\R_y)$ places). \qee

Analogous result is true for (direct) $\sup$-sums of r.d.i.

Thus Theorems~\ref{T1}--\ref{T4} are applicable to the constructed lattice
of r.d.i.\ (l.d.i.) of a given LPDO.
\begin{prop}
\label{prop2}
Any first order LPDO $R$ 
is irreducible (i.e.\ it has no nontrivial divisor ideal couples). 
\end{prop}
\proof Let $R= r_1(x,y) \dx + r_2(x,y)\dy + r_3(x,y)$ but nevertheless
we have some r.d.i.\ $|R_1\}$, $\R > |R_1\} >0$. Necessarily
$\ord_{\dx}(|R_1\}_x)=1$, $\ord_{\dy}(|R_1\}_y)=0$ (up to transposition
$x \leftrightarrow y$), so there exist $R_{1,1} = C_0\o \dx +C_1 \in
|R_1\}$, $R_{1,2} = K_1 \in |R_1\}$, $C_i \in \QY$, $K_1 \in \QX$. Since
$\R \subset |R_1\}$, $R_{1,3} = R_{1,1} - \frac{C_0}{r_1(x,y)} R = K_2
\in |R_1\}$, $K_2 \in \QY$. Due to Corollary~\ref{cor1} of Lemma~\ref{le5}
we have $|R_1\}=0=\QXY$.
\begin{prop}
\label{prop3}
Any LODO $M \in \Q(x)[\dx]$ irreducible in this ring is irreducible as an
element of $\QXY$.
\end{prop}
\proof Suppose we have a nontrivial divisor ideal couple $\L M\R$ so for some
$L \in \L$, $R \in \R$ we have \r{dd2}. Since $M \in \L$, $ M \in \R$ we may
suppose (subtracting a suitable multiple of $M$) that $\ord_{\dx}
(L)=\ord(\PX(L))
< \ord_{\dx}(M) =m$, $\ord_{\dx}(R) < m$. Forming the quotient skew field
$\Q(x,y,\dx,\dy)$ we can write in this skew field $X = Y \o
Q \o L\1$ $\Rightarrow$ $X \o M = Y \o Q \o L\1 \o M = Y \o R$
$\Rightarrow$ $Q\o L\1 \o M =R$. If we will find $\ll(\PX(Q),\PX(L))=
Z = \=Q\o L = \=L \o Q_x \in \QyX$ 
then $\ord_{\dx}(\=L) \leq \ord_{\dx}(L) < m$.
For some $C \in \QY$ we get $C \o \=Q = \wt Q \in \QXY$, $C \o \=L =
\wt L \in \QXY$, $\ord_{\dx}(\wt L) = \ord_{\dx}(\=L) <m$. Then
$\wt L \o Q \o L\1 \o M = \wt L \o R $ $\Leftrightarrow$
$\wt Q \o L \o L\1 \o M = \wt Q \o M = \wt L \o R$. Let us take
$\lcof_{\dy}(\wt L \o R) := \lcof (\PY (\wt L \o R)) =
\lcof_{\dy}(\wt L) \cdot \lcof_{\dy}(R)= C_{\wt L} \o C_R \in \QX$,
$\ord(C_{\wt L}) <m$, $\ord(C_{R}) <m$,
$\lcof_{\dy}(\wt Q \o M) = \lcof_{\dy} (\wt Q) \o M = C_{\wt Q} \o M$,
$\ord(M) =m$. Finally we have $C_{\wt L} \o C_R = C_{\wt Q} \o M$
in $\QX$ which contradicts to the fact that $M$ is irreducible,
$\ord(M)=m$ but the orders of the factors in the l.h.s. are ${}<m$.
\qee

Finally we are able to explain the mystery of the example
of  E.Landau \r{land}:
due to Proposition~\ref{prop2} the chain
$|Q\o Q \o P\} > |Q \o P\} > |P\} > 0$ is maximal,
so the only place where our generalized divisors appear is the
interval between 
$|R\o Q\}$ and  $|Q \}$ in $|R\o Q \} > |Q \} > 0$,
namely we insert here the $I= \ll(|Q \},|P \})$.
In terms of projections $\PX(I)$ is generated by second order
(w.r.t.\ $\dx$) operator $L_{31}=
 \big(x^2\dy^2 + x\dx\dy -(x+1)\dy - \dx\big) \o Q$ while $\PX(R \o Q)$
is a third-order operator. So the \p\  $R$ is reducible in our sense 
(and reducible in $\QyX$) and
it has  a nontrivial ``generalized common (right) divisor'' with
$x^2\dy^2 + x\dx\dy -(x+1)\dy - \dx$ (Definition~\ref{def13}).

\section{ Darboux integrability of nonlinear PDEs and factorization
of linearized equations }
\label{di}

    In  the  XIX-th  century  vast  interest    in
finding  exact solutions to partial
 differential equations resulted in the development
of  methods of Lagrange, Monge, Boole and Ampere. 
 G.~Darboux \ci{darboux}
 generalized the method of Monge (known as the method of intermediate
integrals) to obtain the  most  powerful  method  for  exact  integration  of
partial  differential  equations known in the last century.

Recently in a series of papers \ci{anderson,juras,sokolov}
 the Darboux method was cast into a more precise and efficient (although not
completely algorithmic) form. For the case of a 
single second-order nonlinear PDE of the form
\be
 u_{xy}  = f (x, y, u, u_x, u_y )
\ee{uxy}
their idea consisted in linearization of (\ref{uxy}): using substitution
$u(x,y) \rightarrow u(x,y) + \epsilon v(x,y)$ and cancelling terms with
$\epsilon^n$, $n>1$,  we obtain a LPDE
\be
v_{xy} = Av_x + Bv_y + Cv
\ee{vxy}
with coefficients  depending on $x$, $y$, $u$, $u_x$, $u_y$.
Equations of the type (\ref{vxy}) were studied by Laplace, who invented a
method of transformation (called sometimes the Laplace cascade method)
of (\ref{vxy}).
First af all we remark that for the corresponding LPDO
$L=\dx\o\dy -A\cdot\dx -B\cdot \dy -C$,
\be
L= (\dx -B)\o(\dy-A) + H = (\dy -A)\o(\dx-B) + K,
\ee{fl}
where $H= \dx A -AB -C$, $K= \dy B -AB -C$, are called the Laplace invariants
of (\ref{vxy}). So if eihter $H \equiv 0$ or $K \equiv 0$ our second-order LPDO
$L$
factors in the ``usual'' sense and the solutions of (\ref{vxy}) may be found
via quadratures. If both $H$, $K$  vanish, $L$ is a lLCM of two first-order
LPDO. 
If both $H$, $K$ do not vanish, one can apply the two Laplace
transformations: $L \rightarrow L_1$, $L \rightarrow L_{-1}$, usung the
substitutions
\be
 v_1 = (\dy -A)v, \qquad \quad v_{-1} = (\dx -B)v.
\ee{ls}
These (invertible) 
transformations give two new second order LPDO $L_1$, $L_{-1}$
of the same form with different coefficients iff $H\neq 0$ (resp. $K\neq 0$).
In the generic case one obtains two infinite sequences
$$
L \rightarrow L_1  \rightarrow L_2 \rightarrow \cdots,
$$
$$
L \rightarrow L_{-1}  \rightarrow L_{-2} \rightarrow \cdots.
$$
If one of these sequences is finite (i.e.\ the corresponding Laplace invariant
vanishes and the Laplace transform can not be applied once more) then the final
LPDO $L_i$ is trivially factorable.

One shall certainly take into consideration the original equation (\ref{uxy})
performing all the computations of the Laplace invariants and Laplace
transforms (which allows us to express all the mixed derivatives of $u$
via $x$, $y$, $u$ and the non-mixed $u_{x\cdots x}$, $u_{y\cdots y}$).
\begin{th}
{\rm (\ci{anderson,juras,sokolov})}  A second order, scalar, hyperbolic partial differential equation 
(\ref{uxy}) is Darboux integrable  if  and  only  if  both Laplace  sequences 
 are finite.
\end{th}

In \ci{anderson,juras,sokolov} this method was also generalized for the case
of a general second-order nonlinear PDE 
$$F(x,y,u,u_x,u_y,u_{xx},u_{xy}, u_{yy})=0
$$.

What can be said about factorizability of the operator $L$ in our generalized
sense? 
\begin{th}
$L=\dx\o\dy -a(x,y)\dx -b(x,y)\dy -c(x,y)$ has a nontrivial generalized right
divisor ideal iff one of the Laplace sequences is finite.
\end{th}
\begin{th}
$L=\dx\o\dy -a(x,y)\dx -b(x,y)\dy -c(x,y)$ is a lLCM of two generalized right
divisor ideals 
 iff both Laplace sequences are finite.
\end{th}

The detailed proofs will be given elsewhere.

In fact these theorems demonstrate again that the "generalized" factorization
introduced here enjoys the necessary natural properties: it is invariant
w.r.t.\ the differential substitutions (\ref{ls}) (which destroy the "trivial"
factorizations $L= (\dx -B)\o(\dy-A)$
or  $L= (\dy -A)\o(\dx-B)$).

\section{ Conclusion}
\label{conc}

An obvious and important generalization of our definition of reducibility
(or $\sup$-decomposition) of single LPDO would be a proper definition of 
 decomposition of {\em systems} of LPDO \r{sy1}. Actually a
formal generalization should be
 formulated inductively; for example if a system
of 2 equations has indecomposable first equation $L_1 f =0$ then we may try to
find (generalized) divisor \p\ couples for the second LPDO $L_2$ (i.e.\ the
second equation $L_2 f =0$) forming for $M= L_2$ equations \r{ddd}
{\em modulo the left principal ideal} generated by $L_1$.
The problem of zero-divisors (actually any LPDO is zero divisor now since
it always has a multiple which belongs to the ideal $|L_1\rangle$) is
(apparently) solved using the fact that for non-decomposable $L_1$ there are
no  ``LCM-zero divisors'', i.e.\ if $M \not\in |L_1\rangle$, $R \not \in
|L_1\rangle$ then some their right c.m.\ $M\o X = R \o Y	
 \not\in |L_1\rangle$.
In fact  our proofs rely only on absence of ``LCM-zero divisors''.
Certainly this approach deserves  further thorough study in another
publication. Especially interesting is the possibility to apply such a
generalization to the commutative case (factorized Gr\"obner bases).

It would be interesting to compare our definitions of decomposition of ideals
with the known results on decomposition of ideals in non-commutative rings with
the Ore condition (existence of at least one common multiples for every two
elements) \ci{jac2}.

A more challenging generalization is required for treatment of overdetermined
 linear partial differential systems with {\em several} unknown functions
$f_k$.

The algebraic nature of the set of \p s $\{Q\}$ in \r{ddd}, \r{dd2} also is of
interest: we can multiply $Q$ on the left and on the right with arbitrary LPDO,
but addition of different $Q_i$ is doubly ``stratified'': only $Q_i$
which belong to a fixed $R$ or a fixed $L$ may be added. As we have seen in the
proof of Lemma~\ref{le3} each ``stratum'' of $\{Q\}$ is actually some $\R$
(resp. $\L$).

An algorithm of computation of divisor ideals for a given LPDO would be of big
practical interest. As we have explained in Introduction an algorithm for
$\sup$-de\-com\-po\-si\-ti\-ons is much more important for applications
(Theorems~\ref{T2},~\ref{T3}). One possible approach for algorithmization of
$\sup$-de\-com\-po\-si\-tions may mimic 
the methods of \ci{singer2}. For this purpose we
have to generalize the eigenring algorithm of \ci{singer2} to the case of 
skew differential fields of coefficients $\=\Q(x,y,\dx)$ 
with greater (not algebraically closed) constant subfield $\=\Q(\dx)$.
 Another approach to reducibility testing
may rely on possible generalization of estimates of complexity of coefficients
of factors given in \ci{grig1} for commutative coefficient fields to the case of
the ring $\QxY$. These difficult problems are far beyond the scope of this
short communication.

The theorems proved in section~\ref{di} may provide a basis for algorithmic
checking of Darboux integrability of nonlinear PDEs (provided a suitable
factorisation algorithm for corresponding linearized equations with
coefficients depending on solutions of another PDEs will be found).
 Also we may conjecture
that a generalization of the Darboux integrability method to PDEs of higher
order with arbitrary number of independent variables may be given:
such integrability should be related to representation of the corresponding
linearized LPDO as a lLCM of ``first-order'' generalised divisor ideals.

\section{ Acknowledgements}
\label{ack}


The author wishes to express his special gratitude to Prof.\ F.~Schwarz (GMD,
St.~Augustin, Germany) who communicated the informative and stimulating example
\r{land} and provided a copy of \ci{blum}. Many useful discussions
 about lattice
theory with Prof.~B.V.Yakovlev and Dr.~A.M.Kut'in (Krasnoyarsk)
 are thankfully acknowledged.


\begin{thebibliography}{10}

\bibitem{anderson}
{\sc Anderson,~I.M.,  and   Kamran,~N.}
\newblock  The  Variational  Bicomplex  for  Second  Order  Scalar
     Partial Differential Equations in the Plane.
\newblock {\em CRM technical report}, Utah State University, September 1994.

\bibitem{beke}
{\sc Beke, E.}
\newblock Die {Irreducibilit\"at} der homogenen linearen
  Differentialgleichungen.
\newblock {\em Math. Annalen 45\/} (1894), pp.~278--300.

\bibitem{bir}
{\sc Birkhoff, G.}
\newblock Lattice theory.
\newblock  1967, {\em AMS, Providence, R.I.}


\bibitem{bjork}
{\sc Bj\"ork, J.~E.}
\newblock Rings of differential operators.
\newblock  1979, {\em North-Holland.}

\bibitem{blum}
{\sc Blumberg, H.}
\newblock \"Uber algebraische Eigenschaften von linearen homogenen
Differentialausdr\"ucken.
\newblock {\em Diss., G\"ottingen.} 1912.

\bibitem{darboux}
{\sc Darboux, G.}
\newblock Sur les \'equations aux d\'eriv\'ees partielles du second ordre.
\newblock {\em  Ann. Ecole Normale Sup.,} t. VII (1870), pp.~163--173.

\bibitem{grat}
{\sc Gr\"atzer, G.}
\newblock General lattice theory.
\newblock  1978, {\em Akademie-Verlag, Berlin.}

\bibitem{grig1}
{\sc Grigor'ev, D.~Yu.}
\newblock  Complexity of factoring and calculating the
GCD of linear ordinary differential operators,
\newblock  {\em  J. Symbolic Computation 10\/} (1990), pp.~7--37.


\bibitem{grig2}
{\sc Grigor'ev, D.~Yu.}
\newblock  Complexity of  computing the genus of a system of exterior
  differential equations.
\newblock  {\em  Soviet Math. -- Doklady 39\/} (1989), No~3, pp.~432--436.

\bibitem{ire}
{\sc Ireland, K., and Rosen, M.}
\newblock A Classical Introduction to Modern Number Theory.
\newblock  1982, {\em Springer-Verlag.}

\bibitem{jac1}
{\sc Jacobson, N.}
\newblock Lectures in abstract algebra. v.1.
\newblock  1951.

\bibitem{jac2}
{\sc Jacobson, N.}
\newblock The theory of rings.
\newblock  1943. AMS Math. Surveys, N.Y.

\bibitem{jan}
{\sc Janet, M.}
\newblock Le\c cons sur les syst\`emes d'equations aux deriv\'ees
  partielles.
\newblock  1929, {\em Gautier-Villard.}


\bibitem{juras}
{\sc Juras,~M.}
\newblock Generalized  Laplace  invariants  and  classical  
integration  methods
for  second  order  scalar  hyperbolic  partial  differential 
 equations  in  the  plane.
\newblock {\em Proc. Conf. DIFFERENTIAL GEOMETRY AND APPLICATIONS,}
 Aug. 28 - Sept. 1, 1995, Brno, Czech Republic
Masaryk University, Brno 1996, pp.~275--284.

\bibitem{landau}
{\sc Landau, E.}
\newblock Ein Satz \"uber die Zerlegung homogener linearer 
Differentialausdr\"ucke in irreducible Factoren.
\newblock {\em J. f\"ur die reine und angewandte Math. 124\/}
 (1901/1902), pp.~115--120.

\bibitem{loewy1}
{\sc Loewy, A.}
\newblock {\"Uber} reduzible lineare homogene Differentialgleichungen.
\newblock {\em Math. Annalen 56\/} (1903), pp.~549--584.

\bibitem{loewy2}
{\sc Loewy, A.}
\newblock {\"Uber} vollstandig reduzible lineare homogene
  Differentialgleichungen.
\newblock {\em Math.~Annalen 62\/} (1906), pp.~89--117.

\bibitem{olver}
{\sc Olver, P.~J.}
\newblock Applications of Lie groups to differential equations.
\newblock  1986, {\em Springer-Verlag.}


\bibitem{ore1}
{\sc Ore, O.}
\newblock Theory of non-commutative polynomials.
\newblock {\em Annals of Mathematics 34\/} (1933), pp.~480--508.

\bibitem{ore2}
{\sc Ore, O.}
\newblock Linear equations in non-commutative fields.
\newblock {\em Annals of Mathematics 32\/} (1931), pp.~463--477.

\bibitem{reid}
{\sc Reid, G.~J.}
\newblock Algorithms for reducing a system of PDEs to standard form,
determining the dimension of its solution space and calculating its Taylor
series solutions.
\newblock  {\em Euro. J. Appl. Mech. 2\/} (1991), p. 293--318.

\bibitem{riq}
{\sc Riquier, C.}
\newblock Les  syst\`emes d'equations aux deriv\'ees  partielles.
\newblock  1910, {\em Gautier-Villard.}

\bibitem{singer1}
{\sc Singer, M.~F., and Ulmer, F.}
\newblock Galois groups of second and third order linear
differential equations.
\newblock {\em Journal of Symbolic Computation 16\/} (1993), pp.~9--36.

\bibitem{singer2}
{\sc Singer, M.~F.}
\newblock Testing reducibility of linear differential
operators: a group theoretic perspective.
\newblock {\em  Applicable Algebra in Engineering,
Communication and Computing 7\/} (1996), pp.~77--104.

\bibitem{schwarz}
{\sc Schwarz, F.}
\newblock The Riquier-Janet theory and its applications to nonlinear
evolution equations.
\newblock  {\em Physica D,} 11 (1984), p. 243--351.

\bibitem{sokolov}
{\sc Sokolov,~V.V., and Zhiber,~A.V.}
\newblock On the Darboux integrable hyperbolic equations.
\newblock {\em Physics Letters A} 208 (1995), pp.~303--308.

\bibitem{ts96}
{\sc Tsarev, S.~P.}
\newblock An algorithm for complete enumeration of all
factorizations of a linear ordinary differential operator.
\newblock {\em Proceedings of ISSAC'96} (1996), ACM Press, pp.~226--231.

\end{thebibliography}
\end{document}